\documentstyle[12pt]{article}
\begin{document}

\author{A. S. de Castro and A. de Souza Dutra\thanks{%
E-mail: dutra@feg.unesp.br} \\
UNESP-Campus de Guaratinguet\'a-DFQ\\
Av. Dr. Ariberto Pereira da Cunha, 333\\
C.P. 205\\
12516-410 Guaratinguet\'{a} SP Brasil}
\title{{\LARGE Classes of exact wavefunctions for general time-dependent Dirac
Hamiltonians in 1+1 dimensions}}
\maketitle

\begin{abstract}
In this work we construct two classes of exact solutions for the most
general time-dependent Dirac Hamiltonian in 1+1 dimensions. Some problems
regarding to some formal solutions in the literature are discussed. Finally
the existence of a generalized Lewis-Riesenfeld invariant connected with
such solutions is discussed.

\end{abstract}

\newpage

The interest in solving problems involving time-dependent systems has
attracted the attention of physicists since a long time. This happens due to
its applicability for the understanding of many problems in quantum optics,
quantum chemistry and others areas of physics \cite{abdalla}-\cite{ioan}. In
particular we can cite the case of the electromagnetic field intensities in
a Fabry-P\'{e}rot cavity \cite{abdalla}. In fact this kind of problem still
represents a line of investigation which attracts the interest of physicists 
\cite{feng}\cite{ioan}. However, as has been observed recently by Landim and
Guedes \cite{landim}, the most part of these works deal with nonrelativistic
systems. So they tried to bridge this gap, by studying the problem of a
fermion under the presence of a time-dependent Lorentz vector linear
potential in two-dimensional space-time. For this they used the so called
Lewis-Riesenfeld invariant operator, in order to guess the wavefunction
which solves the problem. Unfortunately, they just presented a formal
solution for the problem under question. In fact, if one consider that in
the most part of the works treating the non-relativistic time-dependent
systems, the solution is not formal only for a limited number of
time-dependence for the potential parameters, it is a little bit strange
that in \cite{landim} there is no limitation over those parameters. In view
of these comments, as a first example, we are going to complete the program
initiated by Landim and Guedes, by solving the problem up to the end,
showing that there exists really some restrictions for a complete solution
of this problem. After that, we will show that there exists at least two
classes of solutions for the time-dependent Dirac equation in 1+1
dimensions. This is done considering the most general combination of Lorentz
structures for the potential matrix. Then we show that the problem studied
in \cite{landim} is only a particular case of one of those classes. Finally
we construct the Lewis-Riesenfeld invariant operators which have our
solutions as eigenfunctions.

Let us begin by presenting the Dirac equation in 1+1 dimensions, in the
presence of a time-dependent potential for a fermion of rest mass $m$, 
\begin{equation}
H\,\,\psi \left( q,t\right) =i\,\frac{\partial \psi \left( q,t\right) }{%
\partial t},\,\,H=\alpha \,p+\beta m+V_{V}\left( q,t\right) +\beta
\,V_{S}\left( q,t\right) +\alpha \,\beta \,V_{P}\left( q,t\right) ,
\label{1}
\end{equation}

\noindent where we used $c=\hbar =1$ and $p$ is the momentum operator. $%
\alpha $ and $\beta $ are Hermitian square matrices satisfying the relations 
$\alpha ^{2}=\beta ^{2}=1,\left\{ \alpha ,\beta \right\} =0$. Here we choose 
$\alpha =\sigma _{1}$and $\beta =\sigma _{3}$, where $\sigma _{1}$ and $%
\sigma _{3}$ are the $2\times 2$ standard Pauli matrices. In this way $\psi $
is a two-component spinor. Furthermore the subscripts for the terms of
potential denote their properties under a Lorentz transformation: $V$ for
the time component of the two-vector potential, $S$ and $P$ for the scalar
and pseudoscalar terms, respectively. The absence of the space component of
the two-vector potential is due to the possibility of its elimination
through a gauge-like transformation.

In the particular case considered by Landim and Guedes \cite{landim}, one
has 
\begin{equation}
H=\alpha \,p+\beta m+f\left( t\right) \,q.
\end{equation}

\noindent As suggested by them, we will use as an Ansatz for the solution
the following spinor 
\begin{equation}
\psi =\left( 
\begin{array}{l}
M_{1}\left( t\right) \\ 
M_{2}\left( t\right)
\end{array}
\right) \,e^{i\,\,\eta \left( t\right) \,q},
\end{equation}

\noindent so that the corresponding coupled equations are

\begin{eqnarray}
i\,\dot{M}_{1} &=&\left[ \left( \,\dot{\eta}+f\right) q+m\right]
M_{1}+\,\eta \,M_{2}\,,  \nonumber \\
&& \\
i\,\dot{M}_{2} &=&\left[ \left( \,\dot{\eta}+f\right) q-m\right]
M_{2}+\,\eta \,M_{1},  \nonumber
\end{eqnarray}

\noindent where the dot denotes differentiation with respect to $t$.
Imposing that $\eta \left( t\right) =\,\int^{t}f\left( \lambda \right)
d\lambda $ we eliminate the spatial dependence of the above equations so
obtaining

\begin{eqnarray}
i\,\dot{M}_{1} &=&m\,M_{1}+\,\eta \,M_{2}\,,  \nonumber \\
&& \\
i\,\dot{M}_{2} &=&-m\,M_{2}+\,\eta \,M_{1}.  \nonumber
\end{eqnarray}

\noindent Now we simplify even more the above equations if we perform the
following identifications 
\begin{equation}
M_{1}\equiv G_{1}\,e^{-\,i\,m\,t},\,M_{2}\equiv G_{2}\,e^{i\,m\,t},
\end{equation}

\noindent getting 
\begin{equation}
\dot{G}_{1}=-\,\eta _{1}\,G_{2}\,,\,\,\,\dot{G}_{2}=-\,\eta _{2}\,G_{1}\,,
\end{equation}

\noindent where $\eta _{1}\equiv \eta \,e^{2\,i\,m\,t}$ and $\eta _{2}\equiv
\eta \,e^{-\,2\,i\,m\,t}$. By deriving the each one of the above equations
it is easy to conclude that one gets the following second-order equations 
\begin{equation}
\ddot{G}_{i}-\left( \frac{\dot{\eta}_{i}}{\eta _{i}}\right) \dot{G}%
_{i}+\,\eta ^{2}\,G_{i}=0,\,\,\,\,\,i=1,\,2.  \label{2}
\end{equation}

\noindent At this point it is convenient to rescale the $G_{i}$ functions as 
$G_{i}=\sqrt{\eta _{i}}\,g_{i}$, one gets finally that 
\begin{equation}
\ddot{g}_{i}+\left[ \frac{1}{2}\left( \frac{\ddot{\eta}_{i}}{\eta _{i}}%
\right) -\left( \frac{\dot{\eta}_{i}}{\eta _{i}}\right) ^{2}+\eta
^{2}\right] g_{i}=0.
\end{equation}

Note that in general these last equations are not somewhat straightforward
to solve as asserted in \cite{landim}. However we note that for a time
exponentially decaying force, these equations takes the form a Schroedinger
equation for the Morse potential in the time variable, without the usual
boundary conditions of such kind of equation.

>From now on we will present an extension of the class of time-dependent
relativistic systems with exact solutions. For this purpose we begin with
the complete Hamiltonian (\ref{1}), and make the more general Ansatz 
\begin{equation}
\psi =\left( 
\begin{array}{l}
M_{1}\left( t\right) \,\,\,\,e^{i\,F_{1}\left( q,t\right) } \\ 
M_{2}\left( t\right) \,\,\,\,e^{i\,F_{2}\left( q,t\right) }
\end{array}
\right) .\,  \label{spinor}
\end{equation}

\noindent Thus we obtain

\begin{eqnarray}
i\,\dot{M}_{1} &=&\left[ \dot{F}_{1}+V_{V}+V_{S}+m\right] M_{1}+\left(
F_{2}^{\prime }\,-V_{P}\right) \,e^{i\left( F_{2}-F_{1}\right)
}\,\,\,M_{2}\,,  \nonumber \\
&&  \label{3} \\
i\,\dot{M}_{2} &=&\left[ \dot{F}_{2}+V_{V}-V_{S}-m\right] M_{2}+\left(
F_{1}^{\prime }\,+V_{P}\right) \,e^{-\,i\left( F_{2}-F_{1}\right) }\,M_{1}, 
\nonumber
\end{eqnarray}

\noindent \noindent where the prime denotes differentiation with respect to $%
q$. As by construction $M_{i}$ do not depend on $q$, one sees that it is
mandatory to get rid of such a dependence in the above equations. As a
consequence two classes of solution emerge, corresponding to $F_{1}=F_{2}$
or $F_{1}=-\,F_{2}$.

Let us analyze the class where $F_{1}=F_{2}=F$. In this case the exponential
factor appearing in the equations disappears and, consequently the only way
of getting the off-diagonal terms independent of $q$, is by imposing that 
\begin{equation}
V_{P}=V_{P}\left( t\right) ,\,\,\,F=\theta _{1}\left( t\right) \,q.
\end{equation}

\noindent On the other hand, the effect of this condition over the diagonal
terms results that 
\begin{equation}
V_{S}=V_{S}\left( t\right) ,\,\,\,\,V_{V}=-\,\dot{\theta}_{1}\left( t\right)
\,q+\theta _{2}\left( t\right) .
\end{equation}

\noindent Note that this class allows the treatment of systems that are at
most linear in the spatial coordinate, and includes the system proposed by
Landim and Guedes as a particular case. The general equations to be solved
in this class is now given by

\begin{eqnarray}
i\,\dot{M}_{1} &=&\left( \chi _{+}+m\right) M_{1}+\eta _{-}\,\,M_{2}\,, 
\nonumber \\
&& \\
i\,\dot{M}_{2} &=&\left( \chi _{-}-m\right) M_{2}+\eta _{+}\,\,\,M_{1}, 
\nonumber
\end{eqnarray}

\noindent with $\chi _{\pm }\equiv \theta _{2}\pm V_{S}$ and $\eta _{\pm
}\equiv \theta _{1}\pm V_{P}$. The above equations can now be decoupled
giving

\begin{eqnarray}
i\,\ddot{G}_{1}-\left[ \chi _{+}+\chi _{-}+i\,\left( \frac{\dot{\eta}_{1}}{%
\eta _{1}}\right) \,\right] \dot{G}_{1}-\left[ \dot{\chi}_{+}-\left( \frac{%
\dot{\eta}_{1}}{\eta _{1}}\right) \chi _{+}-i\,\eta _{+}\eta _{-}+i\,\chi
_{+}\chi _{-}\right] G_{1} &=&0,  \nonumber \\
&& \\
i\,\ddot{G}_{2}-\left[ \chi _{+}+\chi _{-}+i\,\left( \frac{\dot{\eta}_{2}}{%
\eta _{2}}\right) \,\right] \dot{G}_{2}-\left[ \dot{\chi}_{-}-\left( \frac{%
\dot{\eta}_{2}}{\eta _{2}}\right) \chi _{-}-i\,\eta _{+}\eta _{-}+i\,\chi
_{+}\chi _{-}\right] G_{2} &=&0,  \nonumber
\end{eqnarray}

\noindent where we made the same definition as above for the $G_{i}$
functions and now $\eta _{1}\equiv \eta _{-}\,\,\,e^{2\,i\,m\,t}$ and $\eta
_{2}\equiv \eta _{+}\,\,\,e^{-\,2\,i\,m\,t}$. It is easy to verify that the
above equations recall that appearing in (\ref{2}) in their particular case.
Once again it is in general not solvable, showing that we have got a formal
solution, so that if one want to solve the problem until the end, one must
to look for particular cases where these equations have explicit solutions.

>From now on we analyze the class where $F_{1}=-\,F_{2}=F$. In this new
class, the exponential factor appearing in the equations holds and,
consequently the only way of getting the off-diagonal terms independent of $%
q $ in the equations (\ref{3}), is by imposing that 
\begin{equation}
F^{\prime }=-\,V_{P}\left( q,t\right) .
\end{equation}

\noindent Now, the condition of spatial independency of the diagonal terms
implies that 
\begin{equation}
V_{V}=V_{V}\left( t\right) ,\,\,V_{S}=V_{S}\left( q,t\right) =\gamma \left(
t\right) -\dot{F}\left( q,t\right) ,
\end{equation}

\noindent where $\gamma \left( t\right) $ is an arbitrary function of the
time. In this case the equation decouple giving simply 
\begin{equation}
i\,\dot{M}_{1}=\left( V_{V}+\gamma +m\right) M_{1},\,\,i\,\dot{M}_{2}=\left(
V_{V}-\gamma -m\right) M_{2.}
\end{equation}

Now it is easy to obtain the solution of these equations. These are given by 
\begin{eqnarray}
M_{1}\left( t\right) &=&M_{1}\left( 0\right) \,e^{-i\,\left\{
m\,t+\int^{t}\left[ V_{V}\left( \lambda \right) +\gamma \left( \lambda
\right) \right] d\lambda \right\} },  \nonumber \\
&& \\
M_{2}\left( t\right) &=&M_{2}\left( 0\right) \,e^{i\,\left\{
m\,t-\int^{t}\left[ V_{V}\left( \lambda \right) -\gamma \left( \lambda
\right) \right] d\lambda \right\} }.  \nonumber
\end{eqnarray}

At this point some comments are in order. From above we easily conclude that
for this second class of solutions, one really obtains arbitrary non-formal
solutions, provided that the remaining integrals can be done. Here it is
important to remark that this new class of systems allows arbitrary
dependence on the spatial and time variables, in contrast to the previous
class of systems.

In what follows, we try to construct an invariant operator of the Lewis and
Riesenfeld type \cite{riesenfeld}, which have the above classes of solutions
as eigenfunctions. It could be used, as done in \cite{landim}, in order to
suggest the form of the spinor presented as an Ansatz in this work. The
invariant obviously must satisfy the equation 
\begin{equation}
\frac{dI}{dt}=\frac{\partial I}{\partial t}-i\,\left[ I,H\right] =0,
\label{4}
\end{equation}

\noindent and can be written as 
\begin{equation}
I={\cal A}\left( t\right) {\cal \,}\,p+{\cal F}\left( q,t\right) ,
\end{equation}

\noindent where ${\cal A}$ and ${\cal F}$ are $2\times 2$ matrices, whose
elements, in order to guarantee the validity of the above equation (\ref{4}%
), must obey the following set of coupled differential equations 
\begin{eqnarray}
{\cal A}_{11} &=&{\cal A}_{22}=const.\,,\,\,{\cal A}_{12}={\cal A}%
_{21}=const.\,,  \nonumber \\
&&  \nonumber \\
\Lambda _{-}^{\prime } &=&-\,{\cal A}_{11}\,V_{P}^{\prime },\,\Sigma _{-}=%
{\cal A}_{12}\,\left( m+V_{S}\right) ,  \nonumber \\
&&  \nonumber \\
\dot{\Sigma}_{+} &=&\Lambda _{+}+\Lambda _{-}-{\cal A}_{11}\,V_{V}^{\prime
}\,,  \nonumber \\
&&  \label{nonlinear} \\
-i\,\dot{\Sigma}_{-} &=&2\,\Lambda _{+}-i\,{\cal A}_{12}\,V_{P}^{\prime }-i\,%
{\cal A}_{11}V_{S}^{\prime }\,,  \nonumber \\
&&  \nonumber \\
i\,\dot{\Lambda}_{+} &=&2\,\Sigma _{-}\,V_{P}+i\,\Sigma _{-}^{\prime }+i\,%
{\cal A}_{12}V_{V}^{\prime }+2\,\Lambda _{-}\left( m+V_{S}\right) , 
\nonumber \\
&&  \nonumber \\
-i\,\dot{\Lambda}_{-} &=&i\,\Sigma _{+}^{\prime }+i\,{\cal A}%
_{11}V_{P}^{\prime }+i\,{\cal A}_{12}V_{S}^{\prime }-2\,\Lambda _{+}\left(
m+V_{S}\right) ,  \nonumber
\end{eqnarray}

\noindent where we defined that $\Lambda _{\pm }\equiv \frac{{\cal F}%
_{12}\pm {\cal F}_{21}}{2}$ and $\Sigma _{\pm }\equiv \frac{{\cal F}_{11}\pm 
{\cal F}_{22}}{2}$. On the other hand, we are looking for invariant
operators which have the Dirac spinors as eigenfunctions, which implies into
further conditions over the elements of the matrices ${\cal A}$ and ${\cal F}
$. This lead us to the following conditions if we consider the eigenvalue
equation for the invariant operator acting over the spinor (\ref{spinor})
introduced in this work 
\begin{eqnarray}
{\cal A}_{11}(F_{1}^{\prime }+F_{2}^{\prime })+{\cal F}_{11}+{\cal F}_{22}
&=&\chi _{1}\left( t\right) ,\,  \nonumber \\
&&  \label{constraint} \\
\left( {\cal A}_{12}F_{1}^{\prime }+{\cal F}_{21}\right) \left( {\cal A}%
_{12}F_{2}^{\prime }+{\cal F}_{12}\right) &=&\chi _{2}\left( t\right) , 
\nonumber
\end{eqnarray}

\noindent with the functions $\chi _{1,2}$ being arbitrary functions of
time. It is not difficult to show that for the first class one can obtain a
solution, which includes that proposed in \cite{landim} as a particular
case, given by 
\begin{eqnarray}
{\cal A}_{12} &=&{\cal A}_{21}={\cal F}_{12}={\cal F}_{21}=0,\,\,\,{\cal A}%
_{11}={\cal A}_{22}\equiv {\cal A},\,\,\,\,{\cal F}_{11}={\cal F}_{22}\equiv 
{\cal F},  \nonumber \\
&& \\
V_{S} &=&V_{S}\left( t\right) ,\,\,\,V_{P}=V_{P}\left( t\right) ,\,\,\,V_{V}=%
\frac{\dot{{\cal F}}}{{\cal A}}\,q+\chi \left( t\right) .  \nonumber
\end{eqnarray}

For the second class of solutions however, as can be easily realised from
the above set of coupled nonlinear equations (\ref{nonlinear}) and the
constraints coming from (\ref{constraint}), it is quite difficult to extract
simple solutions. For this reason we were not able to verify if there exist
a relativistic invariant which is linear in the momentum, responsible for
the generalized solutions here presented. In fact, as it is typical of
nonlinear equations, different solutions of it could lead to independent
sets of wavefunctions. Finally it is interesting to make some comments, the
first one being the observation that the extension of the solutions
discussed in this work, by introducing a time-dependent mass is quite
simple, as can be seen due to its appearance combined with time-dependent
potentials $V_{S}$ and $V_{V}$. On the other hand, the case of the so called
Dirac oscillator \cite{moshinsky}is not directly solvable from the second
class of solutions, this happens because when $V_{P}$ is linear in the
coordinate, it implies into the need of a $V_{S}$ quadratic in $q$, in order
to guarantee the exact solution of the form guessed in this work.

\noindent {\bf Acknowledgments:} The authors are grateful to FAPESP\ and
CNPq for partial financial support.

\end{document}